\documentclass[a4paper,12pt]{article}
\pdfoutput=1 

\usepackage{jheppub} 

\usepackage[T1]{fontenc} 
\usepackage{tensor}
\usepackage{mathrsfs}

\newcommand{\beq}{\begin{equation}}
\newcommand{\eeq}{\end{equation}}
\newcommand{\be}{\begin{equation}}
\newcommand{\ee}{\end{equation}}
\newcommand{\bea}{\begin{eqnarray}}
\newcommand{\eea}{\end{eqnarray}}
\newcommand{\ben}{\begin{eqnarray*}}
\newcommand{\een}{\end{eqnarray*}}               
\newcommand{\ba}{\begin{align}}
\newcommand{\ea}{\end{align}}
\newcommand{\bt}{\begin{tabular}}
\newcommand{\et}{\end{tabular}}
\newcommand{\bc}{\begin{center}}
\newcommand{\ec}{\end{center}}
\newcommand{\ax}{\alpha}
\newcommand{\bx}{\beta}

\newcommand{\dx}{\delta}

\newcommand{\cO}{\mathcal{O}}
\newcommand{\cT}{\mathcal{T}}

\newcommand{\cC}{\mathcal{C}}
\newcommand{\cD}{\mathcal{D}}

\newcommand{\cK}{\mathcal{K}}
\newcommand{\cN}{{N}}

\newcommand{\cA}{\mathcal{A}}

\newcommand{\cF}{\mathcal{F}}

\newcommand{\cR}{\mathcal{R}}

\newcommand{\cV}{\mathcal{V}}

\newcommand{\jb}{{\bar\jmath }}

\def\cZ{\mathcal{Z}}

\newcommand{\half}{\tfrac12}

\newcommand{\nn}{\nonumber}

\newcommand{\cref}{{\bf [check ref]}}

\def\cR{\mathcal{R}}
\newcommand{\wh}[1]{{\hat{#1}}}
\newcommand{\wt}[1]{{\tilde{#1}}}

\def\e{\text{e}}
\newcommand{\inds}[1]{\indices{#1}}
\def\tbz{{\scriptscriptstyle{(0)}}}
\def\tbo{{\scriptscriptstyle{(1)}}}
\def\la{\langle}
\def\ra{\rangle}
\def\pd{\partial}
\def\jb{{\bar{\jmath}}}
\def\pdb{{\bar{\partial}}}
\def\Kahler{{K\"ahler }}

\def\upd{\mathrm{d}}

\def\lam{\Lambda}
\def\sig{\Sigma}
\def\tbt{{\scriptscriptstyle{(2)}}}
\def\tbth{{\scriptscriptstyle{(3)}}}

\def\om{\omega}

\def\tbth{{\scriptscriptstyle{(3)}}}

\def\Tr{\text{Tr }}

\def\fr{\frac}

\def\Z0{Z^\tbz}
\def\lm{\ell^{\,6}_{\text{M}}}
\def\fr{\frac}
\def\lab{\label}
\def\Re{\text{Re}\,}
\def\Im{\text{Im}\,}
\newcommand{\tabitem}{~~\llap{\textbullet}~~}
\def\tfr{\tfrac}
\def\cZ{{\tilde \chi}}
\def\ep{\epsilon}
\def\wh{\hat}
\def\tbo{{\scriptscriptstyle{(1)}}}
\def\l{\ell}
\def\lM{\ell_{\rm M}}
\def\w{\wedge}
\def\k{\kappa}

\def\wt{\tilde}
\def\T{\mathcal{T}}

\def\calA{\mathscr{A}}
\def\B{\mathcal{B}}

\title{{\boldmath One-modulus Calabi-Yau fourfold reductions with higher-derivative terms}}

\author[a]{Thomas W.~Grimm,}
\author[a]{Kilian~Mayer,}
\author[b]{and Matthias~Weissenbacher\,}

\affiliation[a]{Institute for Theoretical Physics, Utrecht University,\\
Princetonplein 5, 3584 CE Utrecht, The Netherlands}
\affiliation[b]{Kavli Institute for the Physics and Mathematics of the Universe, University of Tokyo,\\
Kashiwa-no-ha 5-1-5, 277-8583, Japan}

\emailAdd{t.w.grimm@uu.nl}
\emailAdd{k.mayer@uu.nl}
\emailAdd{matthias.weissenbacher@ipmu.jp}

\preprint{IPMU17-0181}

\abstract{In this note we consider M-theory compactified on a warped Calabi--Yau fourfold including the eight-derivative terms in the eleven-dimensional action known in the literature. We dimensionally reduce this theory on geometries with one K\"ahler modulus and determine the resulting three-dimensional K\"ahler potential and complex coordinate. The logarithmic form of the corrections suggests that they might admit a physical interpretation in terms of one-loop corrections to the effective action. Including only the known terms the no-scale condition in three dimensions is broken, but we discuss caveats to this conclusion. In particular, we consider additional new eight-derivative terms 
in eleven dimensions and show that they are strongly constrained by compatibility with the Calabi-Yau threefold reduction. We examine their impact on the Calabi-Yau fourfold  reduction and the restoration of the no-scale property.}

\begin{document} 
\maketitle
\flushbottom

\section{Introduction}\label{sec:intro}
Compactifications of eleven-dimensional supergravity, which is the long wavelength limit of M-theory, has been proven to be a fruitful playground for studying various aspects of both conceptual and phenomenological relevance. Compactifying M-theory on a Calabi--Yau three- or fourfold leads to a theory with $N=2$ supersymmetry in five or three dimensions, respectively. The compactifications of the two-derivative action in eleven dimensions were worked out in \cite{Cadavid:1995bk, Haack:2001jz}. It is, however, well known that eleven-dimensional supergravity receives quantum corrections in the form of higher-derivative operators that are suppressed by six or more powers of the eleven-dimensional Planck length $\l_{\rm M}$. The most prominent of these corrections are terms quartic in the Riemann tensor that also have counterparts in both Type II string theories. These terms are known to give rise to quantum corrections to the moduli space of $N=2$ compactifications of Type II string theories on Calabi-Yau threefolds \cite{Antoniadis:1993ze,Antoniadis:1997eg,Grimm:2017okk}.

Besides their phenomenological significance, higher-derivative corrections in eleven dimensions can play an important role in the physics of black objects. These terms in eleven dimensions can lead to corrections to the macroscopic black hole entropy \cite{Kraus:2005vz, Castro:2007hc,Castro:2007ci, Castro:2008ne}. Phenomenological applications of M-theory compactifications are mostly within the framework of F-theory \cite{Vafa:1996xn}. The physics of F-theory is most conveniently studied in the dual formulation given by M-theory on an elliptically fibered Calabi-Yau manifold, see e.g.~\cite{Denef:2008wq,Weigand:2010wm}. The actual physics is then encoded in the non-trivial elliptic fibration. Quantum corrections arising in M-theory compactifications might then lift to corrections in F-theory using the M-theory to F-theory duality. Among these quantum corrections could possibly be a correction to the three-dimensional K\"ahler potential stemming from the $\hat \cR^4$ terms in eleven dimensions. Even though much work was done in the past few years to identify such a correction a definite answer remained elusive, see \cite{Grimm:2013gma, Grimm:2013bha,Junghans:2014zla, Grimm:2014xva, Grimm:2014efa, Minasian:2015bxa, Grimm:2015mua}. Due to the lacking of a fundamental formulation of F-theory we take its dual formulation in terms of M-theory as a definition of F-theory. It is obvious from this point of view, that before attempting to lift certain corrections to four dimensions one first has to understand the dual three-dimensional configuration reasonably well. 

In this note we aim to elaborate on the existence of a correction to the three-dimensional $N=2$ K\"ahler potential by considering the most simple setup. We consider M-theory including known and conjectured $\l_{\rm M}^{ \,6}$ corrections and reduce them on a Calabi--Yau fourfold with only one modulus, namely the overall volume of the compact manifold. We use the corrected fourfold solution involving fluxes and warping found in \cite{Becker:1996gj, Becker:2001pm,Grimm:2014xva} and derive a three dimensional effective action including the gravity multiplet and one vector multiplet. We show that the resulting action is compatible with three-dimensional $N=2$ supersymmetry.  Upon dualizing the vector multiplet into a chiral multiplet we derive the corrected K\"ahler potential and associated complex 
coordinates. We show that the result breaks the no-scale condition in three dimensions leading to a non-vanishing scalar potential for the overall volume. Our results allow us to make some first observations about the M-theory to F-theory limit. We point out that the corrections are reminiscent of one-loop corrections found in three-dimensional effective theories obtained 
after integrating out massive charged modes arising in a circle compactification from four to three dimensions \cite{deBoer:1996mp, deBoer:1996ck,   deBoer:1997kr,Dorey:1997ij, Dorey:1998kq,Tong:2014era}. 

One important caveat to point out is the fact that the eight-derivative terms in the eleven-dimensional action have not been 
established by a supersymmetrization procedure or conclusively determined by string amplitudes. Therefore, one might worry that our result 
obtained from dimensional reduction of the terms suggest in the literature might change significantly if further new higher-derivative terms are included. To check their influence we therefore examine the dimensional reduction of a basis of potentially relevant 
eight-derivative terms of the form $\hat G^2 \hat \cR^3$, where $\hat G$ is the eleven-dimensional four-form field strength. 
We constrain the coefficients in a general ansatz by demanding compatibility with compactification on a Calabi--Yau threefold. 
We find that a specific combination can potentially cancel the logarithmic correction and restore the no-scale condition. However, 
due to the suggested physical interpretation of the logarithmic correction, we believe that the qualitative features of our result will remain 
unchanged once the actual supersymmetric combination of the $\hat G^2 \hat \cR^3$ terms has been determined.

This paper is organized as follows. In section \ref{sugra+background} we review three-dimensional gauged $\cN=2$ supergravity 
with gauged shift symmetries and also state the dictionary for dualizing vector multiplets into chiral multiplets in three dimensions.
We then summarize the $\lM^{\, 6}$-corrected eleven-dimensional supergravity action and the considered fourfold solution with background fluxes and a non-trivial warp-factor. In section \ref{sec:main} we present our reduction ansatz and perform the dimensional reduction of the higher-derivative terms known in the literature including one K\"ahler structure modulus. From the results of the dimensional reduction to three dimensions we then infer the K\"ahler potential and coordinate describing the $\cN=2$ supersymmetric theory. Section \ref{caveats} contains a general basis of relevant eight-derivative terms and constrains their coefficients using five- and four-dimensional supersymmetry arguments. We discuss that such additional 
terms can have severe consequences for the one-modulus reduction.

\section{Setting up higher-derivative Calabi-Yau fourfold reductions}\lab{sugra+background}

This section collects the relevant material to perform the one-modulus dimensional reduction in 
section \ref{sec:main}. We introduce three-dimensional gauged $N=2$ supergravity, the considered 
eleven-dimensional effective action including higher derivative terms, and the eleven-dimensional 
background solution.

\subsection{Three-dimensional gauged $N=2$ supergravity}\lab{sugra}

In this section we briefly review $N=2$ gauged supergravity in three dimensions. Three-dimensional maximal and non-maximal supergravities and their gaugings were exhaustively discussed in \cite{deWit:2004yr}.  The case which relevant for our setting is an $N=2$ supergravity theory 
with a gauged shift symmetry and was studied, for example, in \cite{Berg:2002es,Grimm:2011tb}. This shift symmetry corresponds to an isometry of the geometry describing the scalar field space of the $N=2$ theory.  We consider three-dimensional $\cN=2$ supergravity coupled to chiral multiplets whose complex scalars are denoted by $N^a$. The gaugings are realized along certain isometries $\wt{X}^{ab}$. We also use a constant embedding tensor denoted by $\Theta_{ab}$. The $N=2$ action then reads
\beq
S^\tbth_{N=2}=\int_{M_3} \tfr{1}{2} R \, \star   1-K_{a \bar b}\,  \cD N^a \w \star \, \cD \bar N^{\bar b}-\tfr{1}{2} \Theta_{ab}\, A^a \w F^b-\big(V_{\cT}+V_{F} \big) \star 1\, ,\lab{chiralN=2}
\eeq
where $K_{a \bar b}=\pd_{N^a} \pd_{\bar N^{\bar b}}K$ is a K\"ahler metric with K\"ahler potential $K$. The gauge covariant derivative $\cD N^a$ is defined by
\beq
\cD N^a=\upd N^a+\Theta_{bc}\, \wt{X}^{ab} \, A^c\, .
\eeq
The scalar potential in \eqref{chiralN=2} is given by
\begin{align}
V_{\cT}&=K^{a \bar b} \, \pd_a \cT \pd_{\bar b} \cT-\cT^2\, ,\lab{ScPot1}\\[0.2cm]
V_F&=\e^K \big(K^{a \bar b} D_a W \overline{D_b W}-4 \lvert W \rvert^2\big)\, ,\nn
\end{align}
where $\cT$ is a real function of the chiral fields $N^a$ which will be given more explicitly below. Note that the vectors entering \eqref{chiralN=2} via a Chern-Simons term is non-dynamical. In \eqref{ScPot1} we introduced the hermitian matrix $K^{a \bar b}=(K^{-1})^{a \bar b}$ and a holomorphic  
superpotential $W$. 
\paragraph{Dualizing the action. }We now split the chiral fields as $N^a=(M^I, \, T_\Lambda)$ to obtain a dual description of the action \eqref{chiralN=2} by dualizing the chiral multiplets with bosonic component $T_\lam$ into vector multiplets. The detailed procedure can be found in \cite{Berg:2002es,Grimm:2011tb}, we will therefore just quote the result. Since the dualization is in general not possible, one has to assume that the action \eqref{chiralN=2} is invariant under shifts of $\Im T_\lam$. The relevant gauging is achieved by choosing a constant embedding tensor and 
\be
\wt{X}^{\lam \sig}=-2 i \, \dx^{\lam \sig}\, , \qquad \wt{X}^{I  J}=\wt X^{I \bar J}=0\,, \qquad \wt{X}^{\lam I}
=0\, , \qquad \Theta_{IJ}=0\, .
\ee 
The dual action is then given by
\begin{align}
S^\tbth_{N=2,\,  \rm {dual}}&=\int_{M_3} \tfr{1}{2}R \star 1-\wt K_{M^I \bar M^J} \, \cD M^I \w \star \cD \bar M^{\bar J}+\tfr{1}{4} \wt K_{L^\lam L^\sig} \, \upd L^\lam \w \star\,  \upd L^\sig\nn\\
&+\int_{M_3}\tfr{1}{4} \wt K_{L^\lam L^\sig} \, F^\lam \w \star\,  F^\sig+\tfr{1}{2} \Theta_{\lam \sig} A^\lam \w F^\sig+F^\lam \w \Im \big[ \wt K_{L^\lam M^I} \, \cD M^I \big]\,\nn \\
&-\int_{M_3} \big(V_{\cT}+V_F \big) \star \, 1\, .\lab{dualN=2}
\end{align}
The physical couplings in \eqref{dualN=2} such as $\wt K_{L^\lam L^\sig}=\pd_{L^\lam} \pd_{L^\sig} \wt K$ are now derived from a kinetic potential $\wt K$, which is defined in terms of the K\"ahler potential $K$ by a Legendre transformation
\be
K(M, T)=\wt K(M, L)-\Re T_\lam \, L^\lam\, ,\lab{legendre}
\ee
where the real coordinates conjugate to $\Re T_\lam$ are defined by $L^\lam=-2 K_{T_\lam}=-2 \pd_{T_\lam} K$. The scalars $L^\lam$ are now  scalars in (propagating) vector multiplets. Using the Legendre transformation \eqref{legendre} one can derive many expressions relating derivatives of $K$ with derivatives of $\wt K$ among which the most important ones are
\be
K_{T_{\lam} \bar T_{\sig}}=-\tfr{1}{4} \wt K^{L^\lam L^\sig}\, , \qquad \Re T_{\lam}=\wt K_{L^\lam}\, , \qquad \fr{\pd L^\lam}{\pd T_\sig}=\tfr{1}{2} \wt K^{L^\lam L^\sig}\, .
\ee
These relations are extensively used in the explicit dualization procedure. The scalar potential $V_\T$ reads in the vector multiplet variables
\begin{align}
V_\T&=\wt K^{M^I \bar M^{\bar J}} \pd_{M^{\bar I}} \T \,\pd_{\bar M^{\bar J}} \T-\wt K^{L^\lam L^\sig} \pd_{L^\lam}\T \, \pd_{ L^\sig}\T-\T^2\, ,\nn\\[0.2cm]
\T&=-\tfr{1}{2} L^\lam \, \Theta_{\lam \sig}\,  L^\sig\, .\lab{Texpr}
\end{align}
The scalar potential $V_F$ is in the vector multiplet language given by
\be
V_F=\e^K \, \Big[ \wt K^{M^I \bar M^{\bar J}} D_{M^I} W \, \overline{D_{M^J} W}-\big(4+ L^\sig \, \wt K_{L^\lam L_\sig}\, L^\lam \big)\lvert W \rvert^2\Big]\, .
\ee
The only relevant part to compute the scalar potential in the chiral multiplet formulation \eqref{ScPot1} will be the function $\T$ given in \eqref{Texpr} and the K\"ahler potential $K$, as we will assume a constant superpotential later.

\subsection{Higher derivative corrections in M-theory}
In the following we will review eight-derivative terms in eleven dimensions available in the literature \cite{Gross:1986iv,Duff:1995wd,Green:1997di,Russo:1997mk,Tseytlin:2000sf,Peeters:2005tb,Policastro:2006vt,Hyakutake:2007sm,Liu:2013dna}
that will be relevant for the computation of the corrected three-dimensional, two-derivative effective action arising upon compactification on a Calabi--Yau fourfold $Y_4$. These corrections are given by two sectors. The well known, purely gravitational eight-derivative terms $\wh \cR^4$ are supplemented by terms involving the four-form field strength. The bosonic part of the classical two-derivative $\cN=1$ action in eleven dimensions is given by
\be
2 \k_{11}^2 \, S_{11}=\int_{M_{11}} \hat R \, \hat \ast \, 1-\fr{1}{2} \hat G \w \hat \ast \, \hat G-\fr{1}{6} \hat C \w \hat G \w \hat G\, .\lab{eq:S0}
\ee
These terms are supplemented by certain eight-derivative couplings, such as the famous $\wh \cR^4$ terms 
\be
2 \k_{11}^2 \, S_{\wh \cR^4}=\int_{M_{11}}\big(\hat t_8 \hat t_8-\tfr{1}{24} \epsilon_{11} \epsilon_{11} \big)\hat \cR^4\, \hat \ast \, 1-3^2 2^{13}\,\hat C \w \hat X_8 \,\lab{R4}
\ee
which are related to the R-symmetry and conformal anomaly of the world-volume theory of a stack of $N$ M5 branes \cite{Tseytlin:2000sf}. In addition, there are eight-derivative terms containing the four-form field strength. The latter take the schematic form \cite{Liu:2013dna}
\be
2 \k_{11}^2 \, S_{\wh {\mathcal{G}}} =\int_{M_{11}}-\big(\hat t_8 \hat t_8+\tfr{1}{96} \epsilon_{11} \epsilon_{11} \big)\hat G^2 \, \hat R^3\, \hat \ast \, 1+\hat s_{18} \, \big(\hat \nabla \hat G \big)^2  \, \hat R^2\,\hat \ast \, 1+256 \, \hat Z \hat G \w \hat \ast \, \hat G\, .\lab{Gterms}
\ee
The last term in \eqref{Gterms} was argued to be necessary to ensure Type IIA/M-theory duality when considering Calabi--Yau threefold compactifications \cite{Grimm:2017okk}. The detailed form of the higher-derivative couplings in \eqref{R4}-\eqref{Gterms} is relegated to appendix \ref{app0}. The detailed index structure of the terms of the schemaric form $\big( \hat \nabla \hat G\big)^2 \hat R^2$ can be found e.g.~in \cite{Grimm:2014efa} appendix A.

\subsection{Calabi--Yau fourfold solution including higher derivatives}\label{sec:solution}
In this section we review the fourfold solutions including eight-derivative terms studied in \cite{Becker:1996gj, Becker:2001pm,Grimm:2014xva}. The background solution is taken to be an expansion in terms of the dimensionful parameter \footnote{We follow the conventions of \cite{Tseytlin:2000sf}.}
\be \label{def-alpha}
\ax^2=\fr{(4 \pi \, \k_{11}^2)^{\fr{2}{3}}}{(2\pi)^4 \, 3^2 \cdot 2^{13}}\, ,\qquad \qquad 2\k_{11}^2=(2\pi)^5 \, \l^{\,9}_{\rm M} \, ,
\ee
which reduces to the ordinary direct product solution $\mathbb{R}^{1,2} \times Y_4$ without fluxes and warping to lowest order in $\ax$. All terms at and including $\cO(\ax^2)$ are kept, while higher orders are neglected. At higher order both a warp-factor $A^\tbz$ and fluxes are induced. The solution then takes the form
\begin{align}
\la\upd \hat s^2 \ra &= \e^{\ax^2 \, \Phi^\tbt} \Big(\e^{-2 \ax^2 \, A^\tbt} \eta_{\mu \nu} \, \upd x^\mu \upd x^\nu + 2\e^{\ax^2 \, A^\tbt} \, g_{i \jb}\,  \upd z^i \upd \bar z^{\jb}\Big)\, ,\\
\la \hat G \ra&= \ax \, G^\tbo+ {\rm{dvol}}_{\mathbb{R}^{1,2}} \w \upd \big(\e^{-3 \ax^2 \, A^\tbt} \big)\, .
\end{align}
Using this ansatz one can work out the constraints following from the equations of motion. It turns out that the metric $g_{i \jb}$ is given by an expansion
\be
g_{i \jb}=g^\tbz_{i \jb}+\ax^2 \, g^\tbt_{i \jb}\, , \qquad g_{i \jb}^\tbt \sim \pd_i \pdb_{\jb}\,  \ast^\tbz\big(J^\tbz \w J^\tbz \w F_4 \big)\, ,
\ee
where $g^\tbz$ is the lowest order, Ricci-flat Calabi--Yau metric and $J^\tbz$ is its associated K\"ahler form. We furthermore denote with $F_4$ the non-harmonic part of the third Chern form, which will however be irrelevant for the following discussion, as it drops out of all expressions in the effective action. The metric solution also includes an overall Weyl factor $\Phi^\tbt= -\fr{512}{3}\ast^\tbz \big(c_3^\tbz \w J^\tbz \big)$ and a warp-factor $A^\tbt(z, \bar z)$ satisfying the warp-factor equation
\be\lab{eq:warp}
\Delta^\tbz \, \e^{3 \ax^2  A^\tbt} \, \upd \text{vol}^\tbz_{Y_4}+\fr{1}{2} \ax^2 \, G^\tbo \wedge G^\tbo-3^2 2^{13}\, \ax^2 X^\tbz_8=0\, .
\ee
The background value of the four-form field strength is parametrized by the internal flux $G^\tbo \in H^4 (Y_4)$, which is self-dual with respect to the lowest order Calabi--Yau metric, and a piece proportional to the volume form ${\rm{dvol}}_{\mathbb{R}^{1,2}}$ on $\mathbb{R}^{1,2}$.

\section{M-theory on Calabi--Yau fourfolds with higher-derivative corrections}\lab{sec:main}
In this section we perform the dimensional reduction for the case of a Calabi--Yau fourfold $Y_4$ with $h^{1,1}=1$. This simplified setup allows us to determine the corrected K\"ahler potential and infer the broken no-scale property. Intermediate results of the dimensional reduction are deferred to appendix \ref{app1}.

\subsection{Computation of the quantum-corrected K\"ahler potential}
We now perform the dimensional reduction of eleven-dimensional supergravity on the background reviewed in section \ref{sec:solution}. We will do this for the simplified case $h^{1,1}={\rm dim }\,  H^{1,1} (Y_4)=1$. The single K\"ahler modulus is then given by the volume of the Calabi--Yau fourfold $\cV$. In other words, we expand the K\"ahler form in a single $(1,1)$-form $\omega$ as 
\beq
    J = \cV^{\fr{1}{4}} \omega\ , \qquad \frac{1}{4!}\int_{Y_4} \omega^4 =1 \ , 
\eeq 
where we have normalized $\omega$ to avoid cluttering by numerical factors. Note that $\omega$ is harmonic with 
respect to the zeroth-order background metric $g^\tbz_{i \jb}$.

The simplified analysis with $h^{1,1}=1$ comes with two main advantages. Firstly, one can deduce from the warp-factor equation \eqref{eq:warp} the dependence of the warp-factor on the volume. Secondly, the couplings in the three-dimensional effective action are all topological as opposed to the case for general $h^{1,1}$ considered in \cite{Grimm:2014efa,Grimm:2015mua}.  Upon a rescaling of the metric $g^\tbz_{i \jb} \to \cV^{\fr{1}{4}} g^\tbz_{i \bar j}$ the warp-factor equation should scale homogeneously, mapping a solution of the equation to another solution. This implies that $\tilde  A(z, \bar z, \cV)=\cV^{-1} \, A^\tbt$, where $A^\tbt$ is a solution of the warp-factor equation with respect to the metric $g^\tbz_{i \jb}$. We already noted that the correction to the metric $g^\tbt_{i \jb}$ decouples from the effective action since it only contributes total derivatives. In the following we therefore drop the metric correction from any expression. The reduction ansatz for the metric and the M-theory four form field strength is thus
\begin{align}
\upd \hat s^2&=\e^{\ax^2 \Phi} \Big(\e^{-2 \ax^2  \tilde A}\,g_{\mu \nu} \, \upd x^\mu \upd x^\nu+2 \e^{\ax^2 \tilde A} \, \cV^{\fr{1}{4}}\, i \omega_{i \jb}\, \upd z^i \, \upd \bar z^{\jb}  \Big)\, ,\\
\hat G&=\ax \, G^\tbo+ {\rm{dvol}}_{\mathbb{R}^{1,2}} \w \upd_{Y_4} \big(\e^{-3 \ax^2  A} \big)+F \w \om\, ,
\end{align}
where $F=\upd \mathcal A$ is the field strength of a three-dimensional vector from the expansion of $\hat C$ along the harmonic (1,1)-form $\omega$, $\Phi= \cV^{-\fr{3}{4}} \, \Phi^\tbt$ and  $\tilde A=\cV^{-1} \, A^\tbt$. 
Before we continue with the reduction, let us introduce the useful quantities
\bea
\cZ  =(2\pi)^3 \int_{Y_4} c_3 \w \omega\, ,\qquad \mathscr{A}=\frac{1}{4!} \int_{Y_4} \tilde A \, J^4=\frac{1}{4!}  \int_{Y_4} A^\tbt \, \omega^4\, ,
\eea
where $\cZ $ is a constant depending on the topology of $Y_4$, and $\mathscr{A}$ is a constant depending on the 
warp-factor profile and background metric of $Y_4$.

We will now perform the dimensional reduction of the eleven dimensional action including eight derivative couplings of interest. 
The resulting theory is has $\cN=2$ supersymmetry in three dimensions and contains the gravity multiplet, whose bosonic field is the three-dimensional metric $g_{\mu \nu}$, and a vector multiplet formed by the 3D vector and the volume $\cV$ along with their fermionic superpartners. 
Focusing only on the bosonic part of the action, we first use the reduction results in appendix \ref{app1}
 and then perform a Weyl rescaling to Einstein frame in three dimensions. The resulting action for the kinetic terms takes the form 
\begin{align}
 \kappa_{11}^2\, S_{\rm kin}=&\int_{M_3}\Big[\half R\, \star \, 1-\tfr{9}{16} \upd \log \cV \wedge \,  \star \, \upd \log \cV-\cV^{\fr{3}{2}} \, F \wedge \, \star \, F\lab{einst2}\\
 &\quad~+\tfr{9}{2} \ax^2\, \cV^{-1} \, \calA\, \upd \log \cV \wedge \, \star \, \upd \log \cV+216\, \ax^2 \, \cV^{-\fr{3}{4}} \, \cZ  \, \upd \log \cV \wedge \, \star \, \upd \log \cV\nn\\[0.3cm ]
 &\quad ~-6 \, \ax^2 \, \cV^{\fr{1}{2}} \, \calA\, F \wedge \, \star \, F+384 \, \ax^2 \, \cV^{\fr{3}{4}}\, \cZ  \, F \wedge\, \star \, F\Big]\, .\nn
\end{align}
To compare this result with the general action \eqref{dualN=2} we 
first define $L=\cV^{-\fr{3}{4}}-3 \, \ax^2 \, \mathscr{A} \cV^{-\fr{7}{4}}$ to rewrite \eqref{einst2} as
\begin{align}
\kappa_{11}^2 S_{\rm kin}^\tbth&=\int_{M_3} \fr{1}{2} R \star 1-\fr{1}{L^2} \upd L \wedge \star \, \upd L-\fr{1}{L^2} F \wedge \star \, F\label{einst3}\\
&+\int_{M_3}384 \,  \ax^2 \,\cZ  \, \fr{1}{L} F \wedge \star \, F+384 \, \ax^2 \, \cZ  \, \fr{1}{L}\, \upd L \wedge \star \, \upd L\,. \nn 
\end{align}
It is now easily seen that \eqref{einst3} takes the standard form 
\be
S^\tbth_{\rm stand}=\int_{M_3}\fr{1}{2} R \star 1+\fr{1}{4} \wt{G}_{LL}(L)\, \upd L \wedge \star \, \upd L+\fr{1}{4}\wt{G}_{LL}(L)\,  \, F\wedge \star \, F\, ,
\ee
with
\be\label{kinmet}
\wt{G}_{LL}(L)=-\fr{4}{L^2} \Big(1-384 \, \ax^2 \, \wt \chi  \, L \Big)=-\fr{4}{L^2}+1536 \, \ax^2\, \cZ \, \fr{1}{L}\, .
\ee
We can integrate the metric $\wt G_{LL}$ to obtain the kinetic potential $\tilde K(L)$ and coordinate
\begin{align}
\wt{K}&=4 \log L +1536\, \ax^2 \,  \cZ\, L \, \big(\log (L)-1 \big)+4\, ,\lab{Kpot}\\
L&=\cV^{-\fr{3}{4}}-3 \, \ax^2 \, \mathscr{A}\, \cV^{-\fr{7}{4}}\, , \label{resultL}
\end{align}
where we have chosen the integration constants in a convenient way. 

\paragraph{Determining the K\"ahler potential.} We will now dualize the vector multiplet to a chiral multiplet, whose metric derives from a K\"ahler potential. This is achieved by a Legendre transformation of the kinetic potential as outlined in section \ref{sugra}
\be
K=\wt{K}-L \, \Re\, T\, , \qquad \qquad \Re \, T=\pd_L \wt{K}\, .
\ee
One thus derives the K\"ahler potential  $K(T+\bar T)$ to be
\bea \label{Kpot1}
K &=& 4 \log L-1536 \, \ax^2 \, \wt \chi \, L  \nn\\
   &=&-3 \log \Big(\frac{1}{4!}\int_{Y_4} \e^{4 \ax^2 A} J^4 +512\, \ax^2\,(2 \pi)^3\int_{Y_4}c_3 \wedge J \Big)\,,  
\eea
with corresponding coordinate
\bea
\Re T&=& \fr{4}{L}+1536 \ax^2 \, \wt \chi \, \log L \nn \\
         &=&4 \cV^{\fr{3}{4}}+12 \ax^2 \, \cV^{-\fr{1}{4}}\calA\,-1152\,  \ax^2 \, \cZ \, \log \cV \ .\lab{ReTcorrected}
\eea
All quantities in the K\"ahler potential \eqref{Kpot1} now depend on the modulus $\cV$. On a first sight one might be wonder about that the unusual correction to the K\"ahler coordinate $\propto \log \cV$. We will comment on the physical interpretation of this correction in section \ref{loop}.

Let us stress that the analysis of \cite{Grimm:2015mua} also lead to a K\"ahler potential of the form \eqref{Kpot1}. However, in the analysis performed there, it was not possible to fix all the coefficients in $K$ unambiguously. Furthermore, the discussion of the K\"ahler coordinates was incomplete, due to the presence of many non-topological terms in the more moduli case. In this one modulus analysis we were able to avoid these problems. 
Nevertheless, it is important to point out that the pure warping result of \cite{Grimm:2015mua}, which was inspired by \cite{Martucci:2014ska}, agrees with our findings 
here.

\paragraph{The no-scale condition and the scalar potential.}
The essential key point of this note is that the no-scale condition in three dimensions is broken once $\lM^{\, 6}$ - suppressed corrections to the K\"ahler potential in \eqref{Kpot1} are taken into account. We can straightforwardly compute
\be
K_T \, K^{T \bar T} \, K_{\bar T}=\fr{K_T^2}{K_{T \bar T}}=4-1536 \, \ax^2 \,(2\pi)^3\, {\cV}^{-1} \int_{Y_4}c_3 \w J~ ,\lab{NoScaleCond}
\ee
which indeed shows that the no-scale condition is broken. We can now also use this result to determine the scalar potential. We first evaluate
\begin{align}
V_\T&=K^{T \bar T} \pd_{T} \cT \, \pd_{\bar T} \T-\T^2\nn\\
&=\big(16 \, \cV^{\fr{3}{2}}+\dots \big)^{-1} \, \left[\fr{1}{2} \Big(\fr{\pd \Re T}{\pd \cV} \Big)^{-1} \fr{\pd \T}{\pd \cV} \right]^2-\T^2=0+ \cO(\ax^3)\, , \nn
\end{align}
where we used
\be
\T(\cV)=-\tfr{1}{2} \ax \, \Theta \, \cV^{-\fr{3}{2}}+\dots~, \qquad \qquad \Theta=\fr{1}{2} \int_{Y_4}\omega \w \omega \w G^\tbo\, .
\ee
This means that the only scalar potential comes from the breaking of the no-scale condition. It enters the effective action via the F-term scalar potential
\be
V_F=\e^K \big(K^{T \bar T}\,D_T W \overline{D_{ T}W}-4 \lvert W \rvert^2 \big)=-1536 \,(2\pi)^3\, \ax^2 \, \fr{\lvert W_0 \rvert^2}{\cV^4}\int_{Y_4} c_3 \w J\, ,\lab{FScPot}
\ee
which has a runaway direction for $\cV \to \infty$ if $\int_{Y_4} c_3 \w J < 0$ \footnote{An example with this property and $h^{1, 1}=1$ is the sextic fourfold. For the sextic one finds $\int_{Y_4} c_3 \w \omega =-420$.}. We assumed a constant superpotential $W_0$ in \eqref{FScPot} which may arise from stabilizing complex structure moduli and inserting their fixed values in the GVW superpotential \cite{Gukov:1999ya}
\be
W=\fr{1}{\lM^{\, 3}}\int_{Y_4}G^\tbo \w \Omega \, , \qquad \Omega \in H^{4,0}(Y_4) \, .
\ee 
The runaway behavior of \eqref{FScPot} for large volume $\cV$ signals an instability of the solution for the case of a non-vanishing $W_0$.
This raises doubts about the validity of the reduction for such a non-vanishing $W_0$ as recently stressed in \cite{Sethi:2017phn}. Let us 
emphasize that the vacuum solution around which we expand was supersymmetric and 
therefore $W_0$ should actually vanish in the vacuum for our analysis to be self-consistent.

\subsection{Comments on loop corrections and M/F-theory duality}\lab{loop}

One of the main motivations to study M-theory compactifications on Calabi--Yau fourfolds is its duality to 4D F-theory models with minimal supersymmetry. Upon compactifying the 4D F-theory action on a circle and taking the F-theory limit of vanishing torus-fiber volume vol$(\mathbb{T}^2) \to 0$, one can infer F-theory data from the matching with the 3D M-theory compactification. 
Since F-theory requires an elliptic fibration with $h^{1,1} > 1$ our analysis does not immediately apply to this case. Nevertheless we will try to give some first interpretation of 
the result \eqref{Kpot1} in the context of this duality. 

Let us assume for a moment that our result is valid beyond the one modulus case. While this is semingly straightforward for the K\"ahler 
potential $K$ it is less obvious how to generalize the complex coordinates $T_I$. A reasonable assumption appears to be that 
the $T_I$ contain a correction proportional to 
\beq
  \tilde \chi_I=(2\pi)^3 \int_{Y_4} c_3 \wedge \omega_I \ ,\qquad I = 1,\ldots, \text{dim}\, H^{1,1}(Y_4)\ ,
\eeq 
multiplied with the volume of some submanifold of $Y_4$.
The 3D $N=2$ theory with this K\"ahler potential $K$ and coordinates $T_I$ can then be thought of as a circle-compactified 4D $N=1$ theory with an infinite tower of  (massive) Kaluza-Klein states. The 3D Wilsonian effective action which is valid below some energy scale $\Lambda_{\rm{cutoff}}$ is then calculated by integrating out the massive fields. These massive fields running in loops can then in general modify the physical couplings in the effective action up to arbitrary order in the diagrammatic loop expansion, unless some non-renormalization theorem comes to the rescue. These loop effects are certainly important when employing M/F-theory duality. In particular, the duality may 
mix classical contributions on one side of the duality with quantum corrections to the effective 
action on the other side of the duality. The inclusion of one-loop corrections turned out to be tracking the Chern-Simons terms through 
the M-theory to F-theory limit in compactifications to three dimensions \cite{Grimm:2011fx,Grimm:2015zea,Corvilain:2017luj} 
and five dimensions \cite{Bonetti:2011mw,Bonetti:2013ela,Grimm:2015zea}.

The simplest 4D setting to start with is a supergravity theory with only the $N=1$ gravity multiplet, i.e.~a pure supergravity theory. 
Considering this theory on the background $\mathbb{R}^{1, 2} \times S^1$ the leading perturbative correction to the 3D K\"ahler coordinate was 
determined in \cite{Tong:2014era}. It was inferred from one-loop determinants of fluctuations of the graviton and 
the gravitino (plus their ghosts) around the aforementioned background. The correction was found to be
\be \label{Tong1loop}
\Re T^{\text{1-loop}}_0=2 \pi^2 {M}^2_{\rm pl} \, R^2+\fr{7}{48} \log \big(M_{\rm pl}^2 R^2 \big)\, , 
\ee
where $R$ is the radius of the $S^1$ and $M_{\rm pl}$ is the 4D Planck's mass. This logarithmic correction to the lowest order complex structure is 
reminiscent of the $\log \cV$ correction to the K\"ahler coordinate from the M-theory reduction given in \eqref{Kpot1}. 
However, it is well-known that F-theory setting will not only lead to pure $N=1$ supergravity, but include some moduli 
fields. These are counted by the Hodge numbers $h^{1,1}$, $h^{3,1}$ and $h^{2,1}$ of $Y_4$. Hence, one would need 
to generalize the analysis of \cite{Tong:2014era} to include further complex fields in 4D. 

We are not aware of a study of such a more general setting. However, perturbative and non-perturbative quantum 
corrections to $N=2$ supersymmetric gauge theories without gravity were studied intensively from various point of views, for example, 
in \cite{deBoer:1996mp, deBoer:1996ck,   deBoer:1997kr,Dorey:1997ij, Dorey:1998kq}. We review here 
the case of having a $U(1)^{N_c}$ gauge theory with $N_f$ flavors labelled by $a=1, \dots , N_f$ with Coulomb 
branch masses $q_j^a L^j$, where $q_i^a$ are the charges under the $i$-th $U(1)$ factor and 
$L^i$ is the real scalar in the $i$-th U(1) vector multiplet. For simplicity we set any further real and complex masses to zero. 
The one-loop corrected K\"ahler coordinates are then determined to be (see e.g.~\cite{deBoer:1997kr})
\be
\Re T_i^{\text{1-loop}}=\fr{1}{e^2} \cC_{ij} L^j+\sum_{a}q_i^a \log \big| q_j^a L^j \big|\, .
~\lab{Tloop}
\ee
The corresponding kinetric 
potential was shown to be of the form 
\beq \label{tildeKi}
  \tilde K(L^i) = \fr{1}{2 e^2} \cC_{ij} L^i L^j+\sum_{a}q_i^a L^i (\log \big|  q_j^a L^j \big| - 1)\ . 
\eeq
The corrected coordinates are those of a 3D one-loop Wilsonian effective action with all the massive flavors integrated out. If one now thinks about these coordinates as coming from a 4D $N=1$ supersymmetric F-theory model compactified on a circle one is led to identify the massive KK-modes on the circle with the massive modes that have been integrated out in order to obtain \eqref{Tloop}. From the M-theory side of the duality the massive states responsible for the loop corrections might admit an interpretation in terms of M2-brane states wrapping curves $\cC_a \in H_2(Y_4)$. These states have a mass $q^a_I L^I$ proportional to 
\beq
     \text{vol}(\cC_a) = \int_{\cC_a} J \ , \qquad q^a_I = \int_{\cC^a} \omega_I\ . 
\eeq
Remarkably, the $L$ found in \eqref{resultL} has the elegant more-moduli generalization 
\beq
    L^I = \frac{v^I}{\cV_{\rm W}} \ ,  \qquad \cV_{\rm W} = \frac{1}{4!}\int e^{3 \alpha^2 A} J^4\ ,
\eeq
where $v^I$ are the expansion coefficients in $J =v^I \omega_I$ and $\cV_{\rm W}$ is the standard warped volume.
As suggested by our result \eqref{resultL} this $L^I$ does not include any higher-derivate corrections. 
 Therefore, their contribution via quantum corrections to the three-dimensional effective action
 looks exactly as in \eqref{tildeKi}. 
 
 Note that the results \eqref{Tloop}, \eqref{tildeKi} are obtained in a theory without
 gravity. Coupling to gravity will lead to a logarithmic terms such that the coupled result 
 is expected to be of the form \footnote{Strictly speaking one has to include more vector multiplets $L^0$, $L^\alpha$ 
 with $e^2$ and $\cC_{ij}$ depending on these fields. $L^0$ is the dual variable to $T_0^{\text{1-loop}}$.}
 \beq \label{KL}
    \tilde K(L^i) = - \log(e^2 - \cC_{ij} L^i L^j) +\sum_{a}q_i^a L^i (\log \big|  q_j^a L^j \big| - 1)\ , 
 \eeq
 which expands to \eqref{tildeKi} for small $\cC_{ij}$. Appropriately combining \eqref{Tong1loop} with \eqref{KL} one 
 indeed finds an immediate resemblance with our reduction result \eqref{Kpot}.
Even though this is not a proof that the correction found in \eqref{Kpot1} is a loop correction in the effective action, 
this reasoning supports this interpretation.

\section{Towards a completion of the  $\hat G^2 \hat{\cR}^3$ sector}\label{caveats}

In this section we consider the possibility of having additional $\hat G^2 \hat \cR^3$  terms in the eleven-dimensional action. 
Unfortunately, as of now the supersymmetric completion of the $\hat G^2 \hat \cR^3$-sector is not 
known (see, however, \cite{Hyakutake:2007sm}). The terms we used in section~\ref{sec:main}  have been lifted from corresponding terms in the Type IIA  effective action, which arise at the level of the five-point functions in the Type IIA superstring. Partial indirect conclusions can be drawn at the level of the six-point function \cite{Liu:2013dna}. However, full results remain absent at the level of the six-point function, and especially at higher order $n$-point functions.
It is thus desirable to discuss  possible extensions of the $\hat G^2 \hat \cR^3$ beyond the known terms. 
In this section we study a complete extension of the eleven-dimensional $\hat G^2 \hat \cR^3$  
sector relevant for our Calabi--Yau fourfold reduction.

Instead of computing string amplitudes we take a more indirect approach here. We construct a complete  basis of eight-derivative terms of the schematic form $\hat G^2 \hat \cR^3$, which can contribute to the kinetic terms of the 3D vectors.
To constrain candidate terms we follow the same strategy as in our previous work \cite{Grimm:2017okk}.  Namely, we derive 
restrictions on the higher-dimensional action by looking at constraints arising from lower-dimensional supersymmetry. 
In others words determine the possible extensions of the eleven-dimensional $\hat G^2 \hat \cR^3$  sector, 
which is compatible with 4D and 5D, $N=2$ supersymmetry upon dimensional reduction. 
It turns out that these arguments are very restrictive and allow us to parametrize the  $\hat G^2 \hat \cR^3$ basis with only five parameters. Moreover, these  $\hat G^2 \hat \cR^3$ terms can be chosen to be consistent with the partially known six-point function results \cite{Liu:2013dna}.\footnote{It would be interesting to study the constrains on this sector  arising from demanding that the new structures arising in the Type IIA effective supergravity theory should vanish at the order of the five-point one-loop string scattering amplitude with two NS-NS two-form field and three graviton vertex operator insertions.} 

The implications for our current work is the observation that  by dimensionally reducing the general five-parameter extension of $\hat G^2 \hat \cR^3$-terms generically can modify the kinetic couplings of the 3D vectors. In section  \ref{CY4newAnsatz} we  perform the dimensional reduction of the $\hat G^2 \hat \cR^3$-extensions to three space-time dimensions on a Calabi--Yau fourfold with arbitrary $h^{1,1}$. We find  that this sector naturally gives rise to the same corrections to  the 3D kinetic terms of the vectors depending on $Z_{m \bar n r \bar s} \propto \epsilon_8\epsilon_8 R_{Y_4}^3$ as found in previous work \cite{Grimm:2017okk}. This result will be presented in section \ref{CY4newAnsatz}.

\subsection{Ansatz for a general basis of relevant $ {\hat G}^2  \mathcal{\hat R}^3$ terms}

Let us next discuss the general form of the relevant terms in the basis of $\hat G^2 \hat \cR^3$.
The $\hat G^2 \hat \cR^3$ terms contributing to the three-dimensional effective action are those, which do not contain any 
Ricci tensors or Ricci scalars as these vanish trivially on a Calabi--Yau manifold. Also contractions in which the two four-form field strengths share less than two common indices can not lead to a kinetic term for the 3D vectors. Taking into account the first Bianchi identity for the Riemann tensor,
 a minimal basis of these terms is given in appendix \ref{basis}. Note that terms of the form $(\hat \nabla \hat G)^2  \hat \cR^2$ never contribute to the two-derivative kinetic term of the vector in the one modulus reduction because of the closure of the K\"ahler form. The general expansion of terms which may contribute in addition to \eqref{Gterms}  to the 3D action is then
\be
2 \kappa_{11}^2 \,  S^{\rm{extra, \,gen}}=\ax^2 \, \int_{M_{11}}\sum_{i=1}^{17} \bx_{i} \, \B_i \, \hat \ast 1 \, ,\lab{extra}
\ee
for some coefficients $\beta_i \in \mathbb{R}$.

To restrict the parameters in  the ansatz \eqref{extra}   we  first take a detour to Calabi--Yau threefold compactifications of M-theory with arbitrary number of K\"{a}hler deformations $h^{1,1}(Y_3)$ in section \ref{CY3checkM}. It is known that the resulting theory will be a 5D ungauged supergravity theory with eight real supercharges (${N}=2$). Compatibility with ${N}=2$ supergravity will then lead to constraints on the parameters of the complete basis of eleven-dimensional $\hat G^2 \hat \cR^3$ terms. 
Furthermore, such novel terms in eleven dimensions will generically contribute $H^2 R^3$ terms  to the 10D Type IIA effective action which we discuss in section \ref{CY3checkIIA}, where $H$ is the NS-NS three-form field strength. We compute the imprint of  the the general basis of $\hat G^2 \hat \cR^3$ terms by circular dimensional reduction on $\mathbb{R}^{1,9} \times S^1$ to the ten-dimensional IIA effective action. Analogously to the 5D argument we then perform a dimensional reduction on a Calabi--Yau threefold and compare the findings to 4D, ${N}=2$ supersymmetry,  analogously to  \cite{Grimm:2017okk}.

\subsection{Constraints from IIA on Calabi--Yau threefolds }
\lab{CY3checkIIA}

In this section we first reduce the basis of seventeen $\hat G^2 \hat \cR^3$-terms in the M-theory effective action to 
ten-dimensional Type IIA supergravity on $\mathbb{R}^{1,9} \times S^1$. 
The only terms relevant for us are the ones which arise from
\be
\hat G_{11 M N O} =  e^{\frac{\phi}{3}}H_{M N O}  \;\; ,
\ee
where $M,N,O = 1,\dots,10$, and $11$ denotes the direction along $S^1$.  We then proceed by dimensionally reducing the ten-dimensional novel eight-derivative $H^2 R^3$ terms contributing to the Type IIA supergravity action to four dimensions on a Calabi--Yau threefold. The zero-mode expansion of the NS-NS two-form field gives rise to  $h^{1,1}(Y_3)$ scalars $b^I$ 
\be
B_2= b^I \omega_I \;\; .
\ee
 To be compatible with 4D, ${N} = 2$ supersymmetry the novel $H^2 R^3$ terms cannot modify the kinetic terms of the $b$-scalars \cite{Grimm:2017okk}. Note that in four dimensions the $b$-scalars  combine into the complexified \Kahler moduli $t^a=b^a+i v^a$, where $v^a$ are the $h^{1,1}(Y_3)$ K\"{a}hler moduli.  Furthermore, the kinetic couplings 
arise from a  prepotential $f(t^a)$ given by
\beq\label{preppot}
f(t)=\frac{1}{3!}\cK_{abc}\, t^a t^b t^c-i \, \fr{\zeta(3)}{2 (2\pi)^3} \, \chi(Y_3)\,  .
\eeq
 and are thus heavily constrained. Note that $\chi(Y_3)$ denotes the Euler--characteristic of the Calabi--Yau threefold.
This computation is tedious, we therefore only give the constraints we find
\be
\begin{array}{cclccl}
            \bx_1&=&c_1  \, ,           &   \qquad   \qquad     \bx_2&=&\tfr{1}{8}(c_3-c_4)   \, ,              \\
            \bx_{3}&=&\tfr{1}{16} c_3  \, ,     &                
          \qquad \qquad   \bx_4&=&-2c_2  \, ,                 \\
             \bx_{5}&=&c_2  \, ,                 & \qquad  \qquad     \bx_{6}&=&\tfr{1}{6}(-2c_1+24c_2-c_3)      \, ,                 \\
            \bx_7&=&-12 c_2 + \tfr{1}{12}c_3       \, ,           &    \qquad \qquad      \bx_{8}&=&   \tfr{1}{4}(-2 c_3+c_4-4c_5)  \, ,            \\       
                \bx_{9}&=&\tfr{1}{4}c_3    \, ,             &
        \qquad \qquad       \bx_{10}&=&c_5   \, ,       \\   
                       \bx_{11}&=&-\tfr{1}{4}c_3       \, ,                 &     \qquad \qquad         \bx_{12}&=&\tfr{1}{4}c_3  \, .             \\
                \bx_{13}&=&\tfr{1}{2}c_3   \, ,            &\qquad  \qquad           \bx_{14}&=&c_5       \, ,        \\
                \bx_{15}&=&-\tfr{1}{24} (144 c_2 - c_3 + c_4) \, ,         & \qquad \qquad 
                                \bx_{16}&=&\tfr{1}{4}(-2 c_3+c_4) \, ,      \\
                                         \bx_{17}&=&\tfr{1}{32}(c_3-c_4)       \, ,                 &                 \\
\end{array}\lab{parameters}
\ee
where the five real cofficients $c_i$ parametrize the solution to the constraints. It is remarkable that only five of the original 17 parameters $\beta_i$ remain independent.

One may furthermore wish to check the compatibility of the novel induced $H^2 R^3$ terms making use of the Type IIA--Heterotic duality. Compactifying Type IIA on $K3$ is dual to the Heterotic string on $\mathbb{T}^4$. For our purpose it is enough to show that when compactifying the novel  $H^2 R^3$-terms on $K3$ those do not induce any correction to the 6D action, in particular  no four-derivative terms, which results in one further constraint on the parameters \footnote{The constraint arises from imposing the vanishing of the four-derivative terms $ \chi(K3) \; H^{\scriptscriptstyle{(6D)}}{}^{\mu \nu \rho} H^{\scriptscriptstyle{(6D)}}{}_{\mu}{}^{ \nu_1 \rho_1} R^{\scriptscriptstyle{(6D)}}{}_{\mu \mu_1 \nu  \nu_1} $, with $\mu,\nu =1,\dots,6$.}
\beq
c_3 = c_4 \;\; .
\eeq
This concludes that by tuning one further parameter the proposed maximal extension of $\hat G^2 \hat \cR^3$-terms in the M-theory effective action is fully consistent with the indirect six-point functions results discussed in  \cite{Liu:2013dna}.

\subsection{Constraints from M-theory on Calabi--Yau threefolds }
\lab{CY3checkM}

Similarly one can obtain constraints on the coefficients $\beta_{i}$ by demanding compatibility with ${N}=2$ supersymmetry in 5D upon compactification on a Calabi--Yau threefold. This non-renormalization of the vector multiplets in 5D, ${N}=2$ supergravity can immediately be understood from its description in terms of a real prepotential $\mathcal{F}(X^I)$ and real special coordinates $X^I$. The physical scalars in the vector multiplets have to obey the relation
\be
\cF(X^I)=\tfr{1}{3!} C_{IJK} \, X^I X^J X^K=1\, ,
\ee
where $C_{IJK}$ is a totally symmetric, constant tensor. The latter tensor is in turn determined by the $U(1)$ Chern-Simons terms 
$\sim C_{IJK} \, A^I F^J F^K \subset \mathscr{L}^{\rm{(5D)}}$, which do not receive $\l_{\rm M}^{\, 6}$--corrections. Therefore both the tensor $C_{IJK}$ and also the physical scalars $X^I$ remain uncorrected at $\cO(\lm)$. 

In order to obtain the constraints from ${N}=2$ supergravity in 5D we dimensionally reduce the action \eqref{extra} with general coefficients $\beta_i$ on a Calabi--Yau threefold $Y_3$ to five dimensions. We are interested in the kinetic terms for the vectors and therefore expand
\be
\hat G= F^I \w \om^{Y_3}_I\, ,
\ee
where $F^I$ are the 5D vectors in vector multiplets, $\om_{I}^{Y_3} \in H^{1,1}(Y_3)$ form a basis of harmonic (1,\,1)-forms on $Y_3$, and $I=1, \dots, \text{dim} \,H^{1, 1}(Y_3)$. The constraint imposed by supersymmetry is then that upon reducing \eqref{extra} and applying Schouten identities on $Y_3$ the terms \eqref{extra} do not contribute to the 5D couplings. Since these constraints are expected to be very similar to the ones in Type IIA, we do not state them here but instead work with the ones found in the previous section.

\subsection{The $\hat G^2 \hat {\cR}^3$-extension  on Calabi--Yau fourfolds }\lab{CY4newAnsatz}

Let us next discuss the implications of the maximal  $\hat G^2 \hat {\cR}^3$-extension to the M-theory effective action  on the kinetic couplings of the vectors arising in Calabi--Yau fourfold compactifications. We have found in the previous sections \ref{CY3checkM}  and \ref{CY3checkIIA}  that the  $\hat G^2 \hat {\cR}^3$-extension can be parametrized by four independent coefficients $\bx_1,\bx_2,\bx_3$ and $\bx_5$.
To derive the kinetic terms of the 3D vector one expands the four-form field strength as
\be
\hat G= F^I \w \om_I\, ,
\ee
where $F^I$ are the 3D vectors in vector multiplets, $\om_{I} \in H^{1,1}(Y_4)$ form a basis of harmonic (1,\,1)-forms on $Y_4$, and $I=1, \dots, \text{dim} \,H^{1, 1}(Y_4)$. 
Then the $\hat G^2 \hat {\cR}^3$-extension reduced on a  Calabi--Yau fourfold results in
\be\label{newVecterms}
2 \kappa_{11}^2  \Delta S^{\tbth}_{\rm{kin}}=\int_{M_3}  \, F^I \wedge \star F^J\,\int_{Y_4}  \Big[ 2 i \, \big(144c_2+c_3 \big)\, Z_{m \bar n} \omega_I^{ \bar n m} \omega_{J r}{}^r  + 4 \, c_3\,   Z_{m \bar n r \bar s}\, \omega_I^{ \bar n m} \omega_{J }^{\bar s r}  \Big]  \ast_8 1\;\;\;
\ee
provided we impose $c_4=0$. For the precise definition of the rank four tensor $Z_{m \bar n r \bar s} = \big(\epsilon_8 \epsilon_8 R_{Y_4}^3\big)_{m \bar n r \bar s}$ and $Z_{m \bar n }  = Z_{m \bar n r }{}^r$
 see \cite{Grimm:2014efa}. The structures \eqref{newVecterms} have been found to arise in the kinetic term from the known $\hat G^2 \hat {\cR}^3$-terms in eleven dimensions \cite{Grimm:2014efa}. 
It is intriguing that the 4D, 5D and 6D  constraints on the general $\hat G^2 \hat {\cR}^3$-extension lead to restrictions which naturally give rise to these familiar structures when compactified on a fourfold.

Since the focus of this work was on the one-modulus Calabi--Yau fourfold reduction it is interesting to restrict \eqref{newVecterms}  to this case.
One then finds a shift in the correction to the kinetic term of the 3D vector
\be
2 \kappa_{11}^2 \, \Delta S^{\tbth}_{\rm{kin}}=-\int_{M_3} 144 \, c_2 \, \cV^{-\fr{1}{4}}\, \wt \chi \, F \wedge \star F\, .
\ee
Clearly this implies that one can choose $c_2$ to cancel the corrections found in section \ref{sec:main}. This would restore 
the no-scale property and remove the logarithmic corrections to the coordinates. We do not expect, however, that this is the case, 
due to the physical evidence presented in section \ref{loop}. 

Let us conclude that the extensive checks we have performed in this section allow in principle for a $\hat G^2 \hat {\cR}^3$-extension of the M-theory effective action, which is  fully  consistent with other results known in the literature. However, a string scattering interpretation at the level of the six-point function or a Noether coupling procedure involving the $\hat G^2 \hat {\cR}^3$-sector would be desirable to clarify the 
existence of new terms. However, let us stress that our checks pose the tightest constraints currently known on the completeness of the $\hat G^2 \hat {\cR}^3$-terms. We will discuss the implications of the possible $\hat G^2 \hat {\cR}^3$-extension to the kinetic terms of the vectors \eqref{newVecterms}  in Calabi--Yau fourfold compactifications with arbitrary $h^{1,1}$ in a forthcoming work.


\section{Conclusions}

In this work we extended the study of three-dimensional two-derivative $N=2$ effective actions arising Calabi--Yau fourfold compactifications when including  $\lM^{\, 6}$--higher-derivative terms in eleven-dimensions. Such higher-derivative terms are generally 
needed when considering background fluxes, since only then a non-trivial solution to the eleven-dimensional 
equations of motion and tadpole constraints exists. While at leading order including only a Calabi-Yau metric, the $\lM^{\, 6}$--corrected 
solution does admit a non-trivial warp-factor as well as other higher-derivative corrections. It is a non-trivial task to perturb around this 
solution and derive the resulting three-dimensional effective action. 

To find a fully explicit result, we focused in this work on the special case of a fourfold with one K\"ahler modulus $h^{1,1}=1$ and frozen complex structure moduli. Hence, we derived the three-dimensional effective action for the volume modulus and the vector arising from expanding the M-theory three-form along the normalized K\"ahler form. Dimensionally reducing the known higher-derivative terms and  
matching with $N=2$ supergravity in three dimensions made it necessary to introduce a corrected K\"ahler potential \eqref{Kpot1} and complex coordinate \eqref{ReTcorrected}. Remarkably, we find for the first time the explicit form of the K\"ahler coordinates and show that 
it contains a logarithmic correction in the fourfold volume. Using this result it was easy to show that this system of K\"ahler potential and coordinate breaks the no-scale condition at $\cO(\lM^{\, 6})$. 

The logarithmic correction to the complex coordinate suggests that it might capture a one-loop correction as we discussed in detail in section \ref{loop}. In particular, viewing the three-dimensional effective action as arising from a four-dimensional $N=1$ F-theory effective action on a circle, the logarithmic corrections are expected from integrating out heavy M2-branes wrapped on curves. In the one-modulus reduction we have found that the scalar in the vector multiplet is precisely a curve volume multiplied by the warped volume of the Calabi-Yau fourfold. Furthermore, we have seen that the kinetric potential has a form that is expected when integrating out Kaluza-Klein modes arising from the circle. It is an interesting open task to show that this interpretation is correct by matching the numerical factors of the loop computation with 
the result from topology. 

It is important to stress again that one can be critical with our findings of section \ref{sec:main} referring to the possibility that not 
all $\hat G^2 \cR^3$-terms are known in the literature. We therefore critically examined the possibility of having additional unknown terms in the eleven-dimensional action that might be relevant for our discussion. This can be motivated by the fact that the eleven-dimensional eight-derivative terms are lifted versions of Type IIA terms, which are only tested at the level of the five-point function. 
To draw general conclusions we constructed a basis of potentially relevant eleven-dimensional terms and imposed constraints required by compatibility with four- and five-dimensional $N=2$ supergravity. The constraints we found turn out to be surprisingly severe. Nevertheless 
we have shown that the remaining terms can crucially influence the reduction of section \ref{sec:main} and potentially restore the no-scale condition. While our physical interpretation suggests that the general form of the complex coordinates is indeed correct, one can hope that 
a detailed study of the more-moduli space allows to gain more insights for the precise numerical factors.

~{}\\

\noindent{\bf Acknowledgements.}

\noindent K.M. and M.W. would like to thank the Max-Planck-Intitut f\"ur Physik in Munich for hospitality, where parts of this work were carried out. M.W. was supported by the WPI program of Japan.

\vspace{1cm}
\appendix
\noindent
{\Large \textbf{Appendices}}

\section{Higher-derivative terms}\lab{app0}
In this appendix we collect the explicit expressions for the higher derivative terms in eleven dimensions relevant in the main part.
\begin{equation}\label{R4terms}
S_{R^4}=\frac{1}{2 \kappa^2_{11}} \int_{{M}_{11}} \Big(\wh{t}_8\wh{t}_8-\frac{1}{24} \epsilon_{\scriptscriptstyle 11} \epsilon_{\scriptscriptstyle 11}\Big){\wh{R}}^4 \wh{\ast} 1. 
\end{equation}
The two quantities $\wh{t}_8\, \wh{t}_8 {\wh{R}}^4$ and $\epsilon_{11}\, \epsilon_{11}{\wh{R}}^4$ in \eqref{R4terms} have the index representation
\begin{align}
\wh{t}_8\,\wh{t}_8 {\wh{R}}^4&={{t}_8}\inds{^{M_1 \dotsb M_8}}{t_8}\inds{_{N_1 \dotsb N_8}}{\wh{R}}\inds{^{N_1 N_2}_{M_1 M_2}} \dotsb{\wh{R}}\inds{^{N_7 N_8}_{M_7 M_8}} \, ,\label{Mtht8}\\[0.4cm]
\epsilon_{11}\epsilon_{11}{\wh{R}}^4&={\epsilon_{11}}\inds{^{R_1 R_2 R_3 M_1 \dotsb M_8}}{\epsilon_{11}}\inds{_{R_1 R_2 R_3 N_1 \dotsb N_8}}{\wh{R}}\inds{^{N_1 N_2}_{M_1 M_2}} \dotsb{\wh{R}}\inds{^{N_7 N_8}_{M_7 M_8}}\label{Mthe11}.
\end{align}
The tensor $\wh{t}_8$ is defined as
\begin{align}
\wh{t}_8^{N_1 \dotsb N_8}=&{\tfrac{1}{16}\Big[ -2 \big( \wh{g}^{\scriptscriptstyle{N_1 N_3}}\wh{g}^{\scriptscriptstyle{N_2 N_4}}\wh{g}^{\scriptscriptstyle{N_5 N_7}}\wh{g}^{\scriptscriptstyle{N_6 N_8}}+\wh{g}^{\scriptscriptstyle{N_1 N_5}}\wh{g}^{\scriptscriptstyle{N_2 N_6}} \wh{g}^{\scriptscriptstyle{N_3 N_7}} \wh{g}^{\scriptscriptstyle{N_4 N_8}}+\wh{g}^{\scriptscriptstyle{N_1 N_7}}\wh{g}^{\scriptscriptstyle{N_2 N_8}}\wh{g}^{\scriptscriptstyle{N_3 N_5}} \wh{g}^{\scriptscriptstyle{N_4 N_6}}\big)}\nonumber\\[0.2cm]
&+8\big(\wh{g}^{\scriptscriptstyle{N_2 N_3}}\wh{g}^{\scriptscriptstyle{N_4 N_5}} \wh{g}^{\scriptscriptstyle{N_6 N_7}}\wh{g}^{\scriptscriptstyle{N_8 N_1}}+\wh{g}^{\scriptscriptstyle{N_2 N_5}}\wh{g}^{\scriptscriptstyle{N_6 N_3}} \wh{g}^{\scriptscriptstyle{N_4 N_7}} \wh{g}^{\scriptscriptstyle{N_8 N_1}} +\wh{g}^{\scriptscriptstyle{N_2 N_5}}\wh{g}^{\scriptscriptstyle{N_6 A_7}}\wh{g}^{\scriptscriptstyle{N_8 N_3}}\wh{g}^{\scriptscriptstyle{N_4 N_1}}\big)\nonumber\\[0.2cm]
&-(N_1 \leftrightarrow N_2)-(N_3 \leftrightarrow N_4)-(N_5 \leftrightarrow N_6)-(N_7 \leftrightarrow N_8) \Big]\,.\label{10dt8}
\end{align}
These $\cR^4$-terms are furthermore supplemented by another term quartic in the Riemann tensor. This term however also comprises a three form $\wh{C}_3$. This piece of the higher-derivative action then has the form
\begin{equation}
S_{\wh X_8}= -\frac{3^2 2^{13}}{2 \kappa_{11}^2} \int_{{M}_{11}}\wh{C}_3 \wedge \wh{X}_8
\end{equation}
where eight form $\wh{X}_8$ is defined as
\begin{equation}\label{C3X8}
\wh{X}_8=\frac{1}{192}\left[ \Tr {\wh{\mathcal{R}}}^4-\frac{1}{4}\left(  \Tr {\wh{\mathcal{R}}}^2 \right)^2   \right]
\end{equation}
which is in terms of the (real) curvature two form
\begin{equation}
\wh{\mathcal{R}}\inds{^{M}_{N}}=\frac{1}{2} {\wh{R}}\inds{^{M}_{N N_1 N_2}} \upd x^{N_1} \wedge \upd x^{N_2}.
\end{equation}
In addition to these quartic Riemann tensor terms it was conjectured in \cite{Liu:2013dna} that the complete $\wh{G}_4$ dependence at $\mathcal{O}(\wh{G}_4^2)$ is captured by introducing 
\begin{align}
\wh{t}_8\, \wh{t}_8\, {\wh{G}}^2\, {\wh{R}}^3=&{\wh{t}_8}^{M_1 \dotsb M_8}\,{\wh{t}_8}{}_{N_1 \dotsb N_8}{\wh{G}}{}^{N_1}{}_{M_1 R_1 R_2}{\wh{G}}{}^{N_2}{}_{M_2}{}^{R_1 R_2}{\wh{R}}\inds{^{N_3 N_4}_{M_3 M_4}}\dotsb{\wh{R}}\inds{^{N_7 N_8}_{M_7 M_8}}\nn\\[0.3cm]
\epsilon_{11}\epsilon_{11}\wh{G}^2\, {R}^3=&{\epsilon_{11}}\inds{^{R M_1 \dotsb M_{10}}}{\epsilon_{11}}\inds{_{R N_1 \dotsb N_{10}}}{\wh{G}}{}^{N_1 N_2}{}_{M_1 M_2}{\wh{G}}{}^{N_3 N_4}{}_{M_3 M_4}{\wh{R}}\inds{^{N_5 N_6}_{M_5 M_6}}\dotsb{\wh{R}}\inds{^{N_9 N_{10}}_{M_9 M_{10}}}\, .\nn
\end{align}

\section{Reduction results}\lab{app1}
In this appendix we collect the intermediate results of the reduction, especially of the higher-derivative terms \footnote{The dimensional reduction was performed using the xAct bundle \cite{Martin-Garcia:2007bqa, Nutma:2013zea, Martin-Garcia}.}.

\noindent{\tabitem Kinetic terms from $\wh{t}_8 \wh{t}_8 \wh{R}^4$.}
\be
\int_{M_{11}}\wh{t}_8 \wh{t}_8 \wh{R}^4 \, \wh{\ast}\, 1= \int_{M_3}1536\,  \chi(Y_4)\, \star \, 1-72\, \cV^{\fr{1}{4}}\, \cZ \,\upd \log \cV \wedge\, \star\, \upd \log \cV
\ee

\noindent{\tabitem Kinetic terms from $\ep_{11} \ep_{11} \wh{R}^4$.}
\begin{align}
-\fr{1}{24}\int_{M_{11}}\ep_{11} \ep_{11} \wh{R}^4 \, \wh{\ast}\, 1&=\int_{M_3}768\, \cV^{\fr{1}{4}}\, R \, \star \, 1+24\, \cV^{\fr{1}{4}}\, \cZ\, \upd \log \cV \wedge\, \star\, \upd \log \cV\nn\\
&+\int_{M_3}1536 \, \chi(Y_4)\star 1
\end{align}

\noindent{\tabitem Kinetic terms from $\wh{t}_8 \wh{t}_8 \wh{R}^3\, \wh{G}_4^2$.}
\be
\int_{M_{11}}\wh{t}_8 \wh{t}_8 \,\wh{R}^3\,  \wh{G}_4^2 \, \wh{\ast}\, 1=1152 \, \int_{M_3}\cV^{-\fr{1}{4}}\, \cZ \,F \wedge \, \star \, F
\ee

\noindent{\tabitem Kinetic terms from $\ep_{11} \ep_{11} \wh{R}^3\, \wh{G}_4^2$.}
\be
\fr{1}{96}\int_{M_{11}}\ep_{11} \ep_{11} \wh{R}^3\, \wh{G}_4^2 \, \wh{\ast}\, 1=-384\,\int_{M_3} \cV^{-\fr{1}{4}}\, \cZ \, F \wedge \, \star \, F
\ee

\noindent{\tabitem Kinetic terms from $\wh{Z}\, \wh{G}_4^2$.}
\be
\int_{M_{11}}256\, \wh{Z}\, \wh{G}_4 \wedge \wh{\ast}\, \wh{G}_4=1024 \int_{M_3}\cV^{-\fr{1}{4}}\, \cZ\,F \wedge \star \,  F\, ,
\ee
where the scalar $\wh{Z}$ is 
\be
\wh{Z}=\frac{1}{12}\big( {\wh{R}}_{M_1 M_2}^{~~~~~~~M_3 M_4} {\wh{R}}_{M_3 M_4}^{~~~~~~~M_5 M_6} {\wh{R}}_{M_5 M_6}^{~~~~~~~M_1 M_2}-2 {\wh{R}}_{M_1~~~M_3}^{~~~M_2~~~M_4}  {\wh{R}}_{M_2~~~M_4}^{~~~M_5~~~M_6} {\wh{R}}_{M_5~~~M_6}^{~~~M_1~~~M_2}\big)\,.\nn
\ee
\noindent{\tabitem Einstein Hilbert and kinetic term for $\wh{G}_4$.}

\begin{align}
\int_{M_{11}}\!\!\!\! \tfr{1}{2} \wh{R} \, \wh  \ast \, 1&-\tfr{1}{4} \wh G_4 \wedge \wh \ast \wh G_4=\int_{M_3}\tfr{1}{2}\cV\, \e^{3 \ax^2 A} \big(1-384 \, \ax^2 \,\cV^{-\fr{3}{4}} \cZ \big) R \, \star \, 1-\tfr{1}{4} G^\tbo \wedge \ast^\tbz G^\tbo\nn\\
&+\int_{M_3}\tfr{7}{16} \, \cV\, \upd \log \cV \wedge \star \,  \upd \log \cV-\tfr{3}{16}\,\ax^2 \, A^\tbt \, \upd \log \cV \wedge \star \, \upd \log \cV \nn\\
&+\int_{M_3}240 \, \ax^2 \, \cV^{\fr{1}{4}} \,\cZ\,  \upd \log \cV  \wedge \,  \star \,  \upd \log \cV \nn\\
&+\int_{M_3} \cV^{\fr{1}{2}} \, F \wedge \star \, F-3 \, \ax^2\,  \cV^{-\fr{1}{2}}\, A^\tbt\, F \wedge \star \, F+128 \, \ax^2 \, \cV^{-\fr{1}{2}} \, \cZ\, F \wedge \star \, F\nn\\
{}&{}
\end{align}

\noindent{ \tabitem Chern-Simons term.}

\begin{align}
2 \kappa_{11}^2 \, S_{\rm CS}=\int_{M_{11}}-\fr{1}{6} \wh C_3 \wedge \wh G_4 \wedge \wh G_4=\int_{M_3} \, \ax \, \Theta \, \cA \wedge F\, , \qquad \Theta= \fr{1}{2}\int_{Y_4} \omega \wedge \omega \wedge G^\tbo\, .\nn\\
{}&{}
\end{align}

\noindent{ \tabitem Kinetic terms from $\big(\hat \nabla \hat G \big)^2 \hat R^2$.}
\be
\int_{M_{11}}\hat s_{18} \, \big(\hat \nabla \hat G \big)^2  \, \hat R^2\,\hat \ast \, 1=0
\ee

\section{Basis of relevant $\hat G^2 {\mathcal{R}}^3$ terms}\lab{basis}
The basis for the potentially relevant eight-derivative terms involving the four-form field strength is
\begin{align}
\B_1&={{{{G}}}}_{M_5}{}^{M_7M_8M_9} \,{G}_{M_6M_7M_8M_9} \,  {R}_{MM_2}{}^{M_4M_5}\, {R}^{MM_1M_2M_3} \, {R}_{M_1M_3M_4}{}^{M_6}\lab{GRbasis}\\[0.2cm]
\B_2&={G}_{M_4M_6}{}^{M_8M_9} \, G_{M_5M_7M_8M_9}\, R_{MM_2}{}^{M_4M_5}\, R^{MM_1M_2M_3}\, R_{M_1M_3}{}^{M_6M_7}\nn\\[0.2cm]
\B_3&=G_{M_4M_5}{}^{M_8M_9} \, G_{M_6M_7M_8M_9} \,R_{MM_2}{}^{M_4M_5} \,R^{MM_1M_2M_3}\, R_{M_1M_3}{}^{M_6M_7}\nn\\[0.2cm]
\B_4&=G_{M_6M_7M_8M_9} \, G^{M_6M_7M_8M_9}\,R_{MM_2}{}^{M_4M_5}\,R^{MM_1M_2M_3} \, R_{M_1M_4M_3M_5}\nn\\[0.2cm]
\B_5&=G_{M_6M_7M_8M_9}\,  G_4^{M_6M_7M_8M_9}\, R_{M}{}^{M_4}{}_{M_2}{}^{M_5}\, R^{MM_1M_2M_3}\, R_{M_1M_4M_3M_5}\nn\\[0.2cm]
\B_6&=G_{M_5}{}^{M_7M_8M_9}\,  G_{M_6M_7M_8M_9} \, R_{MM_2}{}^{M_4M_5} \, R^{MM_1M_2M_3} \, R_{M_1M_4M_3}{}^{M_6}\nn\\[0.2cm]
\B_7&=G_{M_5}{}^{M_7M_8M_9} \, G_{M_6M_7M_8M_9}\, R_{M}{}^{M_4}{}_{M_2}{}^{M_5}\, R^{MM_1M_2M_3}\, R_{M_1M_4M_3}{}^{M_6}\nn\\[0.2cm]
\B_8&=G_{M_3M_6}{}^{M_8M_9} \, G_{M_5M_7M_8M_9}\, R_{MM_2}{}^{M_4M_5}\, R^{MM_1M_2M_3}\, R_{M_1M_4}{}^{M_6M_7}\nn\\[0.2cm]
\B_9&=G_{M_3M_5}{}^{M_8M_9} \, G_{M_6M_7M_8M_9}\, R_{MM_2}{}^{M_4M_5}\, R^{MM_1M_2M_3} \, R_{M_1M_4}{}^{M_6M_7}\nn\\[0.2cm]
\B_{10}&=G_{M_3M_6}{}^{M_8M_9} \, G_{M_5M_7M_8M_9}\, R_{M}{}^{M_4}{}_{M_2}{}^{M_5}\, R^{MM_1M_2M_3} \, R_{M_1M_4}{}^{M_6M_7}\nn\\[0.2cm]
\B_{11}&=G_{M_3M_5}{}^{M_8M_9}\,  G_{M_6M_7M_8M_9} \, R_{M}{}^{M_4}{}_{M_2}{}^{M_5} \, R^{MM_1M_2M_3}\, R_{M_1M_4}{}^{M_6M_7}\nn\\[0.2cm]
\B_{12}&=G_{M_4M_7}{}^{M_8M_9} \, G_{M_5M_6M_8M_9}\, R_{M}{}^{M_4}{}_{M_2}{}^{M_5} \, R^{MM_1M_2M_3} \, R_{M_1}{}^{M_6}{}_{M_3}{}^{M_7}\nn\\[0.2cm]
\B_{13}&=G_{M_3M_7}{}^{M_8M_9}\,  G_{M_5M_6M_8M_9}\, R_{MM_2}{}^{M_4M_5}\, R^{MM_1M_2M_3} \, R_{M_1}{}^{M_6}{}_{M_4}{}^{M_7}\nn\\[0.2cm]
\B_{14}&=G_{M_3M_7}{}^{M_8M_9} \, G_{M_5M_6M_8M_9} \, R_{M}{}^{M_4}{}_{M_2}{}^{M_5}\, R^{MM_1M_2M_3} \, R_{M_1}{}^{M_6}{}_{M_4}{}^{M_7}\nn\\[0.2cm]
\B_{15}&=G_{M_5}{}^{M_7M_8M_9} \, G_{M_6M_7M_8M_9}\, R_{MM_1M_2}{}^{M_4}\, R^{MM_1M_2M_3} \, R_{M_3}{}^{M_5}{}_{M_4}{}^{M_6}\nn\\[0.2cm]
\B_{16}&=G_{M_4M_6}{}^{M_8M_9} \, G_{M_5M_7M_8M_9}\, R_{MM_1M_2}{}^{M_4} \, R^{MM_1M_2M_3}\, R_{M_3}{}^{M_5M_6M_7}\nn\\[0.2cm]
\B_{17}&=G_{M_4M_6}{}^{M_8M_9} \, G_{M_5M_7M_8M_9} \, R_{MM_1M_2M_3} \, R^{MM_1M_2M_3}\, R^{M_4M_5M_6M_7}.  \nn
\end{align}

\nocite{*}
\bibliographystyle{utcaps}

\bibliography{references}
\end{document}